\documentclass[10pt, twocolumn, aps, prd, amssymb, amsmath, superscriptaddress, eqsecnum, nofootinbib, notitlepage, longbibliography]{revtex4-1}

\usepackage[colorlinks=true, citecolor=cyan, linkcolor=., anchorcolor=.,filecolor=.,menucolor=.,runcolor=.,urlcolor=.,]{hyperref}
\usepackage[english]{babel}
\usepackage{verbatim, braket, setspace}
\usepackage{graphicx,tikz,mathrsfs}
\usepackage{cancel}
\usepackage[normalem]{ulem}

\def\eqref#1{(\textcolor{blue}{\ref{#1}})}

\DeclareMathOperator{\Tr}{Tr}

\newcommand{\be}{\begin{equation}}
\newcommand{\ee}{\end{equation}}

\newcommand{\mcH}{\mathcal{H}}

\newcommand{\msK}{\mathscr{K}}
\newcommand{\vg}{\vec{g}}

\newcommand{\tth}{\text{th}}

\newcommand{\mfa}{\mathfrak{a}}

\newcommand{\mcO}{\mathcal{O}}

\newcommand{\mcN}{\mathcal{N}}
\newcommand{\mcP}{\mathcal{P}}

\newcommand{\co}{\text{co}}

\begin{document}

\title{Thermal quantum gravity condensates in group field theory cosmology}

\author{Mehdi Assanioussi}
\email{mehdi.assanioussi@desy.de}
\affiliation{II. Institute for Theoretical Physics, University of Hamburg, Luruper Chaussee 149, 22761 Hamburg, Germany \\}
\affiliation{Faculty of Physics, University of Warsaw, Pasteura 5, 02-093 Warsaw, Poland}

\author{Isha Kotecha}
\email{isha.kotecha@aei.mpg.de}
\affiliation{Max Planck Institute for Gravitational Physics (Albert Einstein Institute), Am M\"{u}hlenberg 1, 14476 Potsdam-Golm, Germany \\ }
\affiliation{Institut f\"{u}r Physik, Humboldt-Universit\"{a}t zu Berlin, Newtonstra{\ss}e 15, 12489 Berlin, Germany}


\begin{abstract}

The condensate cosmology programme of group field theory has produced several interesting results. The key idea is in the suggestion that a macroscopic homogeneous spacetime can be approximated by a dynamical condensate phase of the underlying microscopic system of an arbitrarily large number of candidate quanta of geometry. In this work, we extend the standard treatments in two ways: by using a class of thermal condensates, the coherent thermal states, which encode statistical fluctuations in quantum geometry; and, by introducing a suitable class of smearing functions as non-singular, well-behaved generalisations for relational clock frames in group field theory. In particular, we investigate an effective relational cosmological dynamics for homogeneous and isotropic spacetimes, extracted from a class of free group field theory models, and subsequently investigate aspects of its late and early times evolution. We find the correct classical limit of Friedmann equations at late times, with a bounce and accelerated expansion at early times. Specifically, we find additional correction terms in the evolution equations corresponding to the statistical contribution of the new thermal condensates in general; and, a higher upper bound on the number of e-folds, even without including any interactions.

\end{abstract}

\maketitle



\section{Introduction}

The ultimate goal of any theory of quantum gravity is to describe the known physics, while also providing novel falsifiable physical predictions on a measurable scale. One of the most important arenas in this respect is cosmology, with features such as singularity resolution and inflation representing crucial checkpoints for any viable model based on an underlying theory of quantum gravity. It is thus important for any candidate theory to find a suitable continuum and semi-classical regime within the full theory, in which standard cosmology can be approximated, up to effective corrections of quantum gravitational origin. In fact in the group field theory (GFT) approach, such a regime has been suggested via a class of condensate phases of the system \cite{Pithis:2019tvp,Oriti:2016acw,Gielen:2016dss}. \

Group field theory is a statistical field theory
defined formally by a partition function,
\be \label{zgft} 
Z_{\rm GFT} = \int [D\varphi D\bar\varphi] \; e^{-S(\varphi, \bar\varphi)}  
\ee 
where the fields $\varphi$ and $\bar\varphi$ are defined over a Lie group base manifold $G$
\be 
\varphi : G \to \mathbb{C} : g \mapsto \varphi(g)
\ee
and $S$ is a generically non-local function of the fields dictating the system's dynamics. The choice of the base manifold we are interested in is $SU(2)^d \times \mathbb{R}$, with $d \geq 3$. This choice is understood as considering a model of discrete gravity associated with $SU(2)^d$ \cite{Oriti:2005mb,Freidel:2005qe,Oriti:2006se,Oriti:2011jm,Oriti:2013jga,Oriti:2013aqa}, coupled to a scalar matter field taking values in $\mathbb{R}$ providing for instance, the possibility of using the matter field $\phi$ to define a relational frame of reference \cite{Li:2017uao,Oriti:2016qtz,Pithis:2019tvp,Oriti:2016acw,Gielen:2016dss}.
The group field is invariant under a diagonal right action of $SU(2)$ on $SU(2)^d$,
\be \label{closure} 
\varphi(\vg,\phi) = \varphi( g_i h, \phi), \ \forall h \in SU(2)  
\ee
where $\vg= (g_i) \in SU(2)^d$ with $i=1,\dots,d$, and $\phi \in \mathbb{R}$. This symmetry encodes the geometric condition of closure of polyhedra with $d$ faces labelled by $SU(2)$ data, associated with the group fields. These polyhedra are the discrete fundamental quanta of geometry building up spacetime, which can then be modelled as a peculiar, background independent\footnote{Background independence in non-perturbative discrete quantum gravity approaches, like group field theory, is in the radical sense of having no spacetime (and related continuum geometric) structures a priori. General relativistic spacetime (modulo effective quantum gravitational corrections) must then emerge from an underlying non-spatiotemporal, thus manifestly background independent, theory of quantum gravity.} quantum many-body system \cite{Oriti:2017twl,Bianchi:2010gc,Kotecha:2019vvn,Oriti:2014yla,Barbieri:1997ks,Baez:1999tk}. In fact, the GFT partition function \eqref{zgft} dynamically generates discrete quantum spacetimes that are labelled 2-complexes (spin foams), with boundary states given by labelled graphs (spin networks) \cite{Reisenberger:2000zc,Oriti:2005mb,Freidel:2005qe,Oriti:2006se,Oriti:2011jm,Oriti:2013jga,Oriti:2013aqa}. Bulk processes and boundary states of group field theories are thus dual to polyhedral complexes, in most commonly studied models based on loopless combinatorics \cite{Oriti:2014yla}.  \

Such a many-body perspective, with the polyhedral quanta of geometry being excitations of the field $\varphi$, has been very useful. It has allowed for tangible explorations of connections of group field theory (which in turn is strictly related to several other approaches, including loop quantum gravity \cite{Ashtekar:2004eh, thiemann_2007, Han:2005km, Bodendorfer:2016uat}, spin foams \cite{Perez:2012wv, rovelli_vidotto_2014}, causal dynamical triangulations \cite{Loll:2019rdj}, tensor models \cite{Gurau:2011xp} and lattice quantum gravity \cite{Hamber:2009mt}) with quantum information theory \cite{Chirco:2017xjb,Chirco:2019dlx,Chirco:2017wgl,Chirco:2017vhs,Anza:2016fix}, and also with quantum statistical mechanics and thermal physics \cite{ThGFT,Kotecha:2019vvn,Chirco:2018fns,Chirco:2019kez,Kotecha:2018gof,Kegeles:2017ems}. It has further allowed for importing ideas and tools from condensed matter theory, which has been crucial for instance in the development of quantum condensate cosmology \cite{Pithis:2019tvp,Oriti:2016acw,Gielen:2016dss}. 

The present work concerns the incorporation of statistical fluctuations in GFT condensate cosmology, directly building on works in \cite{ThGFT,Oriti:2016qtz,deCesare:2016rsf}, and evaluating the consequences of the presence of such fluctuations in the effective cosmological evolution equations. 
To this end, we use the framework of thermal representations for GFT and thermal condensates, i.e.\ coherent thermal states, introduced in \cite{ThGFT}. In general, the notion of thermality in a background independent system is a particularly subtle issue, mainly due to the absence of an absolute notion of time (see for instance \cite{Kotecha:2019vvn}, and references therein). Nevertheless, thermal condensates \cite{ThGFT} can be defined in the present quantum gravitational system using a generalised notion of Gibbs statistical equilibrium and certain quantum many-body techniques \cite{Kotecha:2019vvn,Chirco:2018fns,Kotecha:2018gof}. These condensates are constructed to mathematically include statistical fluctuations in a given (set of) observable(s), not necessarily energy, and without relying on specific geometric interpretations of the associated quantities e.g. generalised temperature, free energy, entropy \cite{ThGFT,Kotecha:2019vvn,Chirco:2018fns,Kotecha:2018gof}. As will be discussed below, the observable of interest here is spatial volume. The resulting state encodes fluctuations in the underlying quantum geometry, with its corresponding generalised temperature being (by construction) a statistical parameter controlling the strength of these fluctuations \cite{ThGFT,Kotecha:2018gof}. We understand the ensuing system as describing a phase of the universe in which not all quanta of geometry have condensed.

In the context of cosmology, a thermal phase seems likely in any reasonable geometrogenesis scenario, in which the universe transitions from a primordial pre-geometric hot thermal phase, to a phase with an approximate notion of continuum and macroscopic geometry (here, encoded in the notion of a condensate \cite{Oriti:2016acw}),  and in general with a leftover thermal cloud of quanta that have not condensed. In other words, here, we understand a pure, zero temperature GFT condensate, that has been used extensively in previous works \cite{Oriti:2016qtz,deCesare:2016rsf,Pithis:2019tvp,Gielen:2016dss}, as describing a suitable macroscopic phase only at very late times of the system's evolution and not throughout. What we are working with instead is an intermediary phase that would be expected to arise in a transition between a hot pre-geometric phase and a pure condensate. Thus, in this work we present a tentative picture of a universe being modelled as a quantum gravitational condensate of elementary quanta of geometry along with a thermal cloud of the same quanta over it, and in which an early time phase dominated by the thermal cloud and a late time phase dominated by the condensate are generated dynamically. \




We work with a free GFT model and thermal condensates with a static (non-dynamical) thermal cloud, to derive effective generalised evolution equations for homogeneous and isotropic cosmology, which include correction terms originating in the underlying quantum gravitational and statistical properties of the system. At late times we recover the correct general relativistic limit, while at early times we get a bounce between contracting and expanding phases, along with a phase of accelerated expansion characterised by an increased number of e-folds compared with previously reported numbers for the same class of free models. \

The article is organised as follows. In section \ref{cond} we present a summary of the construction of thermal representations in GFT associated with generalised Gibbs states, and define coherent thermal states as candidates for thermal quantum gravitational condensates for applying in condensate cosmology. In section \ref{cosmo} we analyse a free GFT model for effective cosmology in presence of thermal fluctuations introduced via coherent thermal states. We start with explicating the choice of the state in \ref{vgibbs}, based on which we derive the GFT effective equations of motion in \ref{eff1}. In sections \ref{clock} and \ref{eff2}, we reformulate the effective dynamics in terms of relational ``clock" functions, implemented as smearing functions along the $\phi$ direction of the base manifold $G$. As we will discuss, this provides a suitable non-singular generalisation of a relational frame used in previous works in terms of the coordinate $\phi$. We further derive the effective generalised Friedmann equations for flat homogeneous and isotropic cosmology in section \ref{eff3}, recover the correct classical general relativistic limit in a late time regime in \ref{lateev}, and characterise the early time evolution through an assessment of singularity resolution and inflation in \ref{earlyev}. Finally, we close with a discussion of various aspects surrounding the inclusion of statistical fluctuations, interactions and its implications at the level of the effective GFT models, and suggest further extensions of this work in section \ref{disc}. 


\section{Thermal quantum gravity condensates} \label{cond}


\subsection{Bosonic group field theory} \label{gft}


The quantum operator group field theory for bosonic\footnote{Bosonic statistics corresponds to an invariance under particle exchanges in a generic many-body wavefunction. In the present context, this corresponds to a graph automorphism under node exchanges, since a generic many-body wavefunction here describes the state of a labelled graph.} quanta is based on the commutation algebra,

\begin{align} \label{fullCRs} [  {\varphi}(\vg,\phi),  {\varphi}^\dag(\vg',\phi')] = \mathbb{I}(\vg,\vg')\delta(\phi-\phi') \end{align} 
with $[ {\varphi}(\vg,\phi), {\varphi}(\vg',\phi')]=[ {\varphi}^\dag(\vg,\phi), {\varphi}^\dag(\vg',\phi')]=0$. Here, $\delta$ is a delta distribution for functions on $\mathbb{R}$, $\mathbb{I}$ is a delta distribution for gauge-invariant functions on  $SU(2)^d$, and we have dropped the hats over operators to simplify our notation. 
The Hilbert space for a single gauge-invariant quantum is
\begin{align} 
\mcH &= L^2(SU(2)^d/SU(2)) \otimes L^2(\mathbb{R}) 
\end{align}
where the quotient by $SU(2)$ ensures gauge invariance of equation \eqref{closure}. This is the state space of a single quantum $d$-polyhedron labelled with a real number $\phi$. 

In order to work with formally well-defined quantities, we smear the operator-valued distributions $ {\varphi}(\vg,\phi)$ with a suitable basis of functions in $\mcH$, 
\be f_{J\alpha}(\vg,\phi) = D_J(\vg) \otimes T_\alpha(\phi)  \ee
where $\{T_\alpha(\phi)\}_{\alpha}$ is any orthonormal basis of complex-valued smooth functions in $L^2(\mathbb{R})$ for the scalar field, labelled by a discrete index $\alpha$, and $\{D_J(\vg)\}_{J}$ is the complete set of orthonormal Wigner functions for gauge-invariant functions on $SU(2)^d$, given by
$ D_{J}(\vg) = \sum_{\vec{n}}C^{\vec{j}}_{\iota, \vec{n}} \prod_{i=1}^d D^{j_i}_{m_in_i}(g_i) \,. $ 
The basis elements are labelled by $J \equiv (\vec{j},\vec{m},\iota)$, with irreducible representations of $SU(2)$ indexed by $j_i  \in \mathbb{N}/2$, internal representation index $m_i \in (-j_i,...,j_i)$ for each $j_i$,  and intertwiners $C^{\vec{j}}_{\iota, \vec{n}}$ (i.e. $SU(2)$ gauge-invariant tensors) indexed by $\iota$. Then the ladder operators in this new basis (labelled by $J,\alpha$) are given by smearing the operators in the original basis (labelled by $\vg,\phi$), that is
\begin{align} \label{a}  
 {a}_{J\alpha} :=   {\varphi}(f_{J\alpha}) &= \int_{SU(2)^d \times \mathbb{R}} d\vg d\phi\; \bar{D}_J(\vg)\bar{T}_\alpha(\phi)   {\varphi}(\vg,\phi) \;,\\
 \label{adag}   {a}^\dag_{J\alpha} :=   {\varphi}^\dag(f_{J\alpha}) &= \int_{SU(2)^d \times \mathbb{R}} d\vg d\phi\; {D}_J(\vg) {T}_\alpha(\phi)   {\varphi}^\dag(\vg,\phi) \,, 
 \end{align}
where $\bar{D}_J(\vg)$ denotes the adjoint matrix to $D_J(\vg)$. Being simply a change of basis, the algebra is unchanged, now taking the form
\be \label{fullCRalpha} [  {a}_{J\alpha},  {a}^\dag_{J'\alpha'}] = \delta_{JJ'} \delta_{\alpha \alpha'} \ee 
and $[  {a}_{J\alpha},  {a}_{J'\alpha'}]=[{a}^\dag_{J\alpha},{a}^\dag_{J'\alpha'}]=0$. The vacuum state $\ket{0}$, which is specified by the annihilation operators,
\be \label{degvac} {a}_{J\alpha}\ket{0} = 0 \;\;\; \forall J,\alpha  \ee 
generates a Fock space $\mcH_F$ (for symmetric, bosonic states), via cyclic actions of the algebra generators $\{  {a}_{J\alpha},   {a}^{\dag}_{J\alpha},  {1}\}$, given by
\be \mcH_F = \bigoplus_{N \geq 0} {\rm sym}\, \left(\mcH^{\otimes N}\right)\, . \ee

Lastly, we recall that a useful class of states in $\mcH_F$ are the coherent states, defined by
\be \label{zerocs} \ket{\sigma} := D_a(\sigma) \ket{0}\ee
where $D_a$ is the displacement operator,
\be \label{dis} D_a(\sigma) = e^{a^\dag(\sigma) - a(\sigma)} \ee
and 
\begin{align} 
a(\sigma) &= \sum_{J,\alpha} \bar{\sigma}_{J\alpha} a_{J\alpha} = \int d\vg d\phi\; \bar{\sigma}(\vg,\phi)   {\varphi}(\vg,\phi) \label{asigma}\\
a^\dagger(\sigma) &= \sum_{J,\alpha} \sigma_{J\alpha} a^\dag_{J\alpha} = \int d\vg d\phi\; \sigma(\vg,\phi)   {\varphi}^\dagger(\vg,\phi) \label{adsigma}
\end{align}
are the smeared operators, for any single-particle wavefunction $\sigma \in \mcH$.


\subsection{Thermal representations} \label{threp}

The operator setup summarised above has been used to show that effective cosmological dynamics, featuring various aspects such as early time bounces, inflation, cyclic phases etc. \cite{Pithis:2019tvp,Oriti:2016acw,Gielen:2016dss}, can be supported by GFT coherent condensates of the form \eqref{zerocs} above (with varying details depending on the dynamical model of course). Particularly in the context of this article, it has been shown that an effective Friedmann dynamics \cite{Oriti:2016qtz,Oriti:2016ueo,deCesare:2016rsf,deCesare:2016axk,Pithis:2016cxg} can be extracted from an underlying GFT dynamics for coherent states \eqref{zerocs}, thus exhibiting the viability of such condensates as suitable quantum gravitational phases for the cosmological sector \cite{Oriti:2016acw}. \

However, as is evident from equation \eqref{zerocs}, this phase is built over a degenerate and unentangled vacuum \eqref{degvac}, characterised by uncorrelated quanta. Furthermore, it does not incorporate statistical fluctuations in the underlying degrees of freedom. Now from a physical point of view, one would expect that thermal fluctuations play a role in the dynamical evolution of physical observables describing any macroscopic system, including a spacetime built from a large number of elementary quanta of geometry. In other words, studying statistical, thermal properties of quantum spacetimes would be valuable \cite{ThGFT,Kotecha:2019vvn,Chirco:2018fns,Chirco:2019kez,Kotecha:2018gof}, especially in strong gravity regimes like for the early-time Planck-scale era in the context of quantum cosmology models. Along these lines, thermal condensates of the form \eqref{cts} have recently been suggested to be a suitable class of states to consider \cite{ThGFT}. \ 

In general, thermal effects are encoded in statistical mixed states, represented by density operators defined on a representation Hilbert space (here, $\mcH_F$), and of particular interest are statistical equilibrium states. In the present background independent context of GFT, these are given by generalised Gibbs states of the form,
\be \rho_{\{\beta_\ell\}} = \frac{1}{Z_{\{\beta_\ell\}}}e^{-\sum_{\ell}\beta_\ell \mcO_\ell} \ee
where, $\{\mcO_\ell\}$ is a finite set of observables of interest and $\{\beta_\ell\}$ is a multivariable generalised inverse temperature \cite{Kotecha:2018gof,Chirco:2018fns,Kotecha:2019vvn}. One can extend this formalism for generalised Gibbs states even further in order to address the question of including thermal fluctuations in condensates, by constructing thermal representations as is done for standard finite-temperature quantum field theories. Such an extension of the GFT formalism to include thermal representations was introduced in \cite{ThGFT}, by using tools from thermofield dynamics \cite{Takahasi:1974zn,Matsumoto:1985mx,Umezawa:1982nv,Umezawa:1993yq,Khanna:2009zz}. Below we review the main details of this construction.

A simple class of generalised Gibbs states that are properly normalisable\footnote{We refer to \cite{Kotecha:2018gof,ThGFT} for the details of proof of normalisation for this class of states.}, and are of interest to us particularly in the present context of cosmology, is 
\be \label{qmrho}  {\rho}_\beta = \frac{1}{Z} e^{-\beta  \mcP} \ee
where $\mcP$ is a positive, extensive operator on $\mcH_F$, and $\beta \in \mathbb{R}_+$. In the context of gravity for instance, $\mcP$ could be chosen to be a spatial volume operator \cite{Kotecha:2018gof,Kotecha:2019vvn}, as will be done later in section \ref{cosmo}. For now, we leave it as this more general class of states, along the lines in \cite{ThGFT}. \

We first define the zero temperature phase ($\beta = \infty$), which is given in terms of the Hilbert space,
\begin{align} \mcH_\infty = \mcH_F \otimes \tilde{\mcH}_F \,. \end{align} 
Here, the Hilbert space $\tilde{\mcH}_F$ is conjugate to $\mcH_F$ under the action of an anti-unitary (modular conjugation or tilde conjugation) operator \cite{Ojima:1981ma,Matsumoto:1985mx,Landsman:1986uw,Celeghini:1998sy,Khanna:2009zz}, and is the Hilbert space of the tilde degrees of freedom. For a detailed discussion on the tilde and non-tilde degrees of freedom, their relations, and possible meanings, we refer to \cite{ThGFT} and references therein.  For now, we note that the zero temperature Hilbert space $\mcH_\infty$ can be understood as describing a bipartite system, with the non-tilde degrees of freedom residing in $\mcH_F$, and its associated algebra generated by $\{a,a^\dag,1\}$, being our original system of interest. Thus also, the observables of interest here are those that belong to the algebra of non-tilde $a$ (or equivalently, $\varphi$) operators.

The space ${\mcH}_\infty$ is a Fock space on the cyclic vacuum 
\begin{align} \ket{0_\infty} = \ket{0} \otimes \ket{\tilde{0}} \end{align} 
with ladder operators $ \{ a_{J,\alpha}, {a}^\dag_{J,\alpha}, \tilde{a}_{J,\alpha},  {\tilde{a}}^\dag_{J,\alpha} \}_{\beta = \infty}$ that satisfy,
\begin{align}\label{ladderOp}
[  {a}_{J\alpha},  {a}^\dag_{J'\alpha'}] = [  {\tilde{a}}_{J\alpha},  {\tilde{a}}^\dag_{J'\alpha'}] = \delta_{JJ'} \delta_{\alpha \alpha'}
\end{align}
and all other commutators, including those between tilde and non-tilde operators, vanish. The vacuum $\ket{0_\infty}$ satisfies
\begin{align}    {a}_{J\alpha}\ket{0_\infty} =    {\tilde{a}}_{J\alpha}\ket{0_\infty} = 0 \;\;\; \forall J,\alpha \,. \end{align} 

The thermal algebra $\{b_{J\alpha},b_{J\alpha}^\dag,\tilde{b}_{J\alpha},\tilde{b}_{J\alpha}^\dag\}_{0 < \beta < \infty}$ is then introduced via thermal Bogoliubov transformations of the generators \eqref{ladderOp}, given by
\begin{align} \label{qmbog1}
b_{J\alpha} &:= \cosh \left[\theta_{J\alpha}(\beta)\right] \, a_{J\alpha} - \sinh \left[\theta_{J\alpha}(\beta)\right] \, \tilde{a}_{J\alpha}^\dag  \\ \label{qmbog2}
\tilde{b}_{J\alpha} &:= \cosh \left[\theta_{J\alpha}(\beta)\right] \, \tilde{a}_{J\alpha} - \sinh \left[\theta_{J\alpha}(\beta)\right] \, {a}_{J\alpha}^\dag 
\end{align}
and analogous expressions for their adjoints $b_{J\alpha}^\dag$ and $\tilde{b}_{J\alpha}^\dag$. The parameters $\theta_{J\alpha}(\beta)$ encode complete information about the statistical state, being functions of $\beta$ and eigenvalues of the observables characterising the state. The temperature-dependent annihilation operators define a new vacuum,
\begin{align} \label{qmbvac} b_{J\alpha}\ket{0_\beta} = \tilde{b}_{J\alpha}\ket{0_\beta} = 0  \;\;\; \forall J,\alpha \end{align} 
called a thermal vacuum, at inverse temperature $\beta$. It is cyclic, in turn generating a thermal Hilbert space $\mcH_\beta$ via the action of the $\beta$-ladder operators, which create and annihilate $b$-quanta over $\ket{0_\beta}$. It is in fact inequivalent to all the different vacua at different temperatures, including the zero temperature $\ket{0_\infty}$ vacuum \cite{ThGFT}. Thus also their corresponding thermal representations labeled by the parameter $\beta$ are all inequivalent, each representing a distinct statistical phase of the system. Notice that $\ket{0_\beta}$ is an entangled state with quantum correlations between pairs of $a$ and $\tilde a$ quanta. 

In the present case, the functions $\theta_{J\alpha}(\beta)$ are uniquely associated with the Gibbs states $\rho_\beta$ of equation \eqref{qmrho}. In practice, they are usually determined by the following condition for the number operator\footnote{In principle, the condition \eqref{req} of identifying observable averages in the two representations, must be satisfied by the full algebra \cite{ThGFT,Takahasi:1974zn,Matsumoto:1985mx,Umezawa:1982nv,Umezawa:1993yq,Khanna:2009zz}.},
\be \label{req} \Tr_{\mcH_F}(\rho_\beta a_{J\alpha}^\dag a_{J\alpha}) = \bra{0_\beta} a_{J\alpha}^\dag a_{J\alpha} \ket{0_\beta}_{\mcH_\beta} \ee
where evaluating the right hand side (using inverse Bogoliubov transformations) gives,
\begin{align} 
\Tr_{\mcH_F}(\rho_\beta a_{J\alpha}^\dag a_{J\alpha}) = \sinh^2\left[\theta_{J\alpha}(\beta)\right]
\end{align}
for all $J,\alpha$. Lastly, Bogoliubov transformations are canonical. Thus, the $\beta$-ladder operators satisfy the same bosonic commutation algebra as before, i.e.
\begin{align} \label{qmbcr1}
[ b_{J\alpha}, b_{J'\alpha'}^\dag] = [ \tilde{b}_{J\alpha}, \tilde{b}_{J'\alpha'}^\dag] = \delta_{JJ'} \delta_{\alpha \alpha'}
\end{align}
and all other commutators vanish. 


\subsection{Coherent thermal states} \label{thcs}

Coherent thermal states are the canonical coherent states over a thermal vacuum $\ket{0_\beta}$ which are invariant under the tilde conjugation \cite{ThGFT,1985JOSAB...2..467B,MANN1989273,mann,Khanna:2009zz}. They are elements of the thermal Hilbert space $\mcH_\beta$, obtained through the action of displacement operators of the form \eqref{dis}, and are defined as
\begin{align} \label{cts} 
\ket{\sigma,\bar{\sigma};\beta} := D_a(\sigma)D_{\tilde{a}}(\bar{\sigma})\ket{0_\beta}  
\end{align} 
where, $D_{\tilde{a}}$ is a displacement operator of the same form as \eqref{dis} but for the $\tilde a$ ladder operators. Coherent thermal states are eigenstates of $\beta$-annihilation operators with temperature-dependent eigenfunctions,
\begin{align}
b_{J\alpha}\ket{\sigma,\bar{\sigma};\beta} = (\cosh\left[\theta_{J\alpha}\right] - \sinh\left[\theta_{J\alpha}\right])\sigma_{J\alpha}\ket{\sigma,\bar{\sigma};\beta} \\
\tilde{b}_{J\alpha}\ket{\sigma,\bar{\sigma};\beta} = (\cosh\left[\theta_{J\alpha}\right] - \sinh\left[\theta_{J\alpha}\right])\bar{\sigma}_{J\alpha}\ket{\sigma,\bar{\sigma};\beta}
\end{align}
where $\sigma_{J\alpha}=(f_{J\alpha} , \sigma)_{\mcH}$. They are not however the eigenstates of the  original annihilation operators $a_{J\alpha}$, belonging to our system of interest. This fact induces the emergence of non-trivial thermal contributions, along with coherence properties, in the expectation values of operators originally defined on $\mcH_\infty$, or its relevant subspace $\mcH_F$. For example, the average number density of the mode ${J\alpha}$ in a coherent thermal state is
\begin{align} \label{expNa}
\bra{\sigma,\bar{\sigma};\beta}a^\dag_{J\alpha}a_{J\alpha}\ket{\sigma,\bar{\sigma};\beta} = |\sigma_{J\alpha}|^2 + \sinh^2\left[\theta_{J\alpha}\right]
\end{align} 
which contains both, the usual coherent number density and an additional thermal contribution. 

It is important to remark here that our use of the basis $f_{J\alpha}$, in particular of the countable basis $\{T_\alpha\}_{\alpha \in \mathbb N}$ for $L^2(\mathbb{R})$, in order to develop the finite-temperature GFT formalism in terms of the ladder operators (equations \eqref{a}-\eqref{adag} and \eqref{qmbog1}-\eqref{qmbog2}), was crucial. The observables that one might consider in a chosen model must admit domains of definition which contain the sector of Hilbert space that one is interested in, here coherent thermal states. Otherwise, no calculation could be carried out without running into divergences and ill-defined expressions. This in particular applies to $\phi$-dependent operators, not smeared with $T_\alpha$. For instance, if one considers the number density operator as a function of $\phi$,
\begin{align} \nonumber
 N_J(\phi) &=\int d\vg  d\vg' \bar{D}_J(\vg') {D}_J(\vg) {\varphi}^\dag(\vg,\phi) {\varphi}(\vg',\phi) \\ &= a_J^\dag(\phi) a_J(\phi) 
\end{align}
then the calculation of the expectation value in a coherent thermal state would give
\begin{align} \label{deltan}
 \langle N_J(\phi) \rangle_{\sigma,\bar{\sigma};\beta} = |\sigma_J(\phi)|^2 + \sinh^2[\theta_J(\phi)]\ \delta(\phi - \phi) 
\end{align}
which is clearly ill-defined due to the presence of the Dirac delta distribution $\delta(0)$ in the thermal part evaluated at the singular point. We thus need to consider smeared observables such as the operator $a^\dag_{J\alpha}a_{J\alpha}$ in \eqref{expNa}, where now the thermal contribution contains a well-defined Kronecker delta $\delta_{\alpha \alpha}$ coefficient instead.

At this point, we have introduced the necessary kinematical aspects of the theory in order to discuss an effective cosmological model incorporating statistical fluctuations of quantum geometry, which is the subject of the rest of the paper. 

\section{Thermal condensate cosmology} \label{cosmo}

We start by presenting the effective free GFT dynamics in a condensate phase with fluctuating geometric volume. We then introduce the notion of a reference clock function and reformulate the setup, including the effective dynamics in terms of functional quantities and equations of motion, with respect to a generic class of these clock functions. Based on this, we present an effective, relational, homogeneous and isotropic cosmological model, and discuss the late and early times evolution it describes. 

\subsection{Condensates with volume fluctuations} \label{vgibbs}

Since we are interested in the homogeneous and isotropic cosmological sector, the main observable of interest is the volume operator, in particular the volume associated to a (spatial) sub-manifold given by a foliation parametrized by a clock function (see section \ref{clock}). Recall that the GFT volume operator associated with a generic many-body state in $\mcH_F$ is   
\be
 V:= \sum_{J,\alpha} v_J\ a^\dagger_{J\alpha} a_{J\alpha}
\ee
where $v_J \in \mathbb{R}_{\geq 0}$ is the volume assigned to a single quantum with a configuration $J = (\vec{j},\vec{m},\iota)$. This is an extensive positive operator on $\mcH_F$, and its action on any multi-particle state gives the total volume by summing up the volume contribution $v_J$ from each quantum. \

Further, we select a statistical state of the form \eqref{qmrho} such that the generator $\mcP$ is the above volume operator, that is a volume Gibbs state \cite{Kotecha:2018gof,Kotecha:2019vvn} of the form
\begin{align} \label{crho}  {\rho}_\beta = \frac{1}{Z_\beta} e^{-\beta  V} \end{align} 
which encodes a statistically fluctuating volume of quantum spacetime. \

The quantum gravity condensate that we are interested in is thus a coherent thermal state of the form \eqref{cts}, but associated specifically with the volume Gibbs state in equation \eqref{crho}. In other words, we are interested in using a state $\ket{\sigma,\bar{\sigma};\beta}$, specified by two functions, the condensate wavefunction $\sigma \in \mcH$, and the Bogoliubov parameter $\theta_{J\alpha}(\beta)$ that is identified by the Bose number distribution of the state \eqref{crho},
\be \label{bose}
\sinh^2\left[ \theta_{J\alpha} (\beta) \right] = \frac{1}{e^{\beta v_J} - 1 } \,.
\ee
Notice that since the spectrum of $V$ is independent of the modes $T_\alpha$, the functions $\theta_{J\alpha}$ are also independent of them. Thus $\theta_{J\alpha} = \theta_J$, and we will drop the labels $\alpha$ in quantities associated with $\theta$ from here on. Also, notice the following important property of our chosen state,
\be \label{zerolim}
\lim_{\beta \to \infty} \ket{\sigma,\bar{\sigma};\beta} = \ket{\sigma,\bar{\sigma}} = D_a(\sigma) D_{\tilde{a}}(\bar{\sigma}) \ket{0,\tilde{0}} 
\ee
thanks to which all results of the previous works in GFT cosmology are reproduced when the fluctuations are turned off completely. \


In the present context of extracting effective cosmological models from a candidate background independent theory of quantum gravity, the use of relational observables is of utmost importance. As mentioned previously, past works in GFT cosmology have interpreted and used the base manifold coordinate $\phi$ as a relational matter clock, and considered quantities such as $N(\phi)$ as relational observables. However, we have also noticed above that in the present setting such quantities (see equation \eqref{deltan}) contain UV divergences related to occurrences of the ill-defined $\delta(\phi=0)$ distributions, which had in turn prompted us to change the basis\footnote{Further details about this aspect and the definition of a basis-independent clock are discussed in section \ref{clock}.} to $T_\alpha$, as a first step in the inclusion of thermal fluctuations in the context of GFT condensates. It follows that in this basis, we are interested in $\alpha$-dependent quantities defined by a partial sum over $J$, such as 
\be V_\alpha = \sum_{J} v_J\ a^\dagger_{J\alpha} a_{J\alpha} 
 \ee
and its statistical average in the thermal condensate (which itself now includes statistical fluctuations in volume),
\be
\langle V_\alpha \rangle_{\sigma,\bar{\sigma};\beta} = \sum_J v_J (|\sigma_{J\alpha}|^2 + \sinh^2\left[ \theta_{J} (\beta) \right]) \,.
\ee

Finally, in the context of the relational dynamics discussed later in section \ref{clock}, we are essentially considering the case of a non-dynamical thermal cloud, with only the condensate part of the full system being dynamical. This is understood as a first approximation of the more general case with dynamical thermal fluctuations. We will return to this point later in sections \ref{earlyev} and \ref{disc}.


\subsection{Effective group field theory dynamics} \label{eff1}

A generic GFT action with a local kinetic term and a non-local interaction term (higher than quadratic order in the fields) takes the form,
\begin{align} \label{action}
 S = \int d\vec{g}d\phi \, \bar{\varphi}(\vec{g},\phi) \msK(\vec{g},\phi) \varphi(\vec{g},\phi) \;+\; S_{\text{int}}[\varphi,\bar{\varphi}] \end{align} 
and gives the following classical equation of motion,
\begin{align} \label{EoMCl} \msK(\vec{g},\phi) \varphi(\vec{g},\phi) \;+\; \frac{\delta S_{\text{int}}[\varphi,\bar{\varphi}]}{\delta \bar{\varphi}(\vec{g},\phi)} = 0 \;. \end{align} 
In the corresponding quantum theory on $\mcH_F$, the equation of motion is,
\begin{align} \msK(\vec{g},\phi) \widehat{\varphi}(\vec{g},\phi) + \widehat{\frac{\delta S_{\text{int}}[\varphi,\bar{\varphi}]}{\delta \bar{\varphi}(\vec{g},\phi)}} = 0 \end{align} 
with some choice of operator ordering (and the hat notation reinstated temporarily). An effective equation of motion can then be derived from the above operator equation by taking its expectation value in a special class of quantum states implementing a notion of semi-classical and continuum approximations\footnote{In line with previous works, we understand the implementation of semi-classical and continuum approximations in the specific sense of using the class of coherent states (which are well-known to be the most classical quantum states, with a peak on a pair of classical conjugate variables), and a condensate phase (described by a collective condensate variable, and with a non-zero order parameter), respectively.}. Here, we take the coherent thermal states introduced above as this class of states. We thus consider,
\begin{align} \label{EoMEff} \bra{\sigma,\bar{\sigma};\beta} \msK(\vec{g},\phi) \widehat{\varphi}(\vec{g},\phi) + \widehat{\frac{\delta S_{\text{int}}[\varphi,\bar{\varphi}]}{\delta \bar{\varphi}(\vec{g},\phi)}} \ket{\sigma,\bar{\sigma};\beta} = 0 \;.\end{align}

As a first step to investigate the role of statistical fluctuations of quantum geometry in condensate cosmology, we focus here only on the free part. This would allow us to display clearly the impact of non-zero thermal fluctuations. In other words, any difference in results that we find, as compared to previous zero temperature free theory studies, could then be attributed directly to the presence of these statistical fluctuations. Therefore, restricting to the free kinetic term, we obtain
\begin{align}
 \bra{\sigma,\bar{\sigma};\beta} \msK(\vec{g},\phi) {\varphi}(\vec{g},\phi) \ket{\sigma,\bar{\sigma};\beta}
= \msK(\vec{g},\phi) \sigma(\vg,\phi)  \,.
\end{align}
Further using the Peter-Weyl decomposition for $\sigma$,
\begin{align}
\sigma(\vg,\phi) &= \sum_{J} D_{J}(\vg) \sigma_J(\phi) \,,
\end{align}
and considering the following kinetic term (which is a standard choice, see for instance \cite{Gielen:2018xph} and references therein),
\begin{align} \label{kin}
\msK = \msK_0(\vec{g}) + \msK_1(\vec{g})\partial_\phi^2
\end{align} 
such that
\begin{align}
\msK_0(\vec{g}) (D_{J}(\vg) \sigma_J(\phi) ) &=  B_JD_{J}(\vg) \sigma_J(\phi) \\ 
\msK_1(\vec{g})\partial_\phi^2 (D_{J}(\vg) \sigma_J(\phi) ) &= A_J D_{J}(\vg)\partial_\phi^2\sigma_J(\phi) 
\end{align}
we obtain the following equations of motion,
\begin{align} \label{EqM}
\partial_\phi^2\sigma_J(\phi) - M_J \sigma_J(\phi) = 0 \,,\quad \forall J 
\end{align} 
where $M_J := - \frac{B_J}{A_J}$. 

Thus, we see that the free GFT dynamical equation of motion in a coherent thermal state is identical to the case where one considers a simple coherent state \eqref{zerocs} in $\mcH_F$, with no thermal cloud. But as we can already anticipate, observable averages (like volume) will have thermal contributions in general, consequently modifying their evolution equations.

This concludes the derivation of the effective GFT equation of motion using a thermal coherent state. However, as we have mentioned earlier, calculations with observables in a $(\vg,\phi)$ basis leads to singularities in the $\phi$-dependent quantities. This brings us to the following section where we address the question of defining and applying a suitable time reference frame (a clock), and offer a preliminary interpretation of the resultant quantities.


\subsection{Smearing functions and reference clocks} \label{clock}

As we have emphasised before, the use of $\phi$ as a reference clock is not possible here since the quantities of interest, like $\braket{V(\phi)}$, are mathematically ill-defined, which prompted us to define quantities like $\braket{V_\alpha}$ instead. Below, we generalise this even further and introduce generic\footnote{While satisfying certain boundary conditions, see equations \eqref{bcond}.} square-integrable, complex-valued smooth functions, 
\be t(\phi) = \sum_{\alpha} t_\alpha T_\alpha(\phi) \ee
in order to define observables and their dynamics as functionals of $t(\phi)$ (which will later be interpreted as relational). This brings us to the aspect of smearing.


In the quantum operator setup summarised in sections \ref{gft} and \ref{threp}, instead of smearing the operators with a set of basis functions $f_{J\alpha}(\vg,\phi)$, we could instead smear the algebra generators with a complete set of more general smearing functions $F(\vg,\phi)$ (usually also satisfying additional analyticity and sufficient decay properties). In particular for the $\phi$ variable, this would amount to smearing with smooth functions, say $t(\phi)$. This would result in an equivalent, but basis-independent algebraic setup, as commonly encountered in Weyl C*-algebraic theory associated with bosonic quanta\footnote{See \cite{Kotecha:2018gof,Kegeles:2017ems} for details of a Weyl algebraic formulation in group field theory.}. For our actual purposes, we retain the use of the Wigner basis $D_J(\vg)$, in order to retain also the associated geometric interpretation of (functions of) the spin labels $J$ as it is standard in both GFT and loop quantum gravity, while in the $\phi$ direction we smear with a function $t(\phi)$. In other words, we are interested in smeared operators of the form, 
\begin{align} 
 a_J(t) &:= \int d\vg d\phi \, \bar{D}_J(\vg) \overline{t}(\phi) {\varphi}(\vg,\phi) \\
 a^\dag_J(t) &:= \int d\vg d\phi \, {D}_J(\vg) {t}(\phi) {\varphi}^\dagger(\vg,\phi) 
\end{align}
which are now understood as \emph{functional} (relational) ladder operators, with respect to the \emph{function} $t(\phi)$. By extension, the observable of interest is the volume operator, which now takes the form,
\be
 V_t := \sum_{J} \ v_J\ a^\dagger_{J}(t) a_{J}(t)
\ee
which is interpreted as the operator associated to a spatial slice labeled by the function $t$. \ 

Notice that in general, $t$-relational operators, and their expectation values, are non-local functions of their $\phi$-relational counterparts. For instance, the average volume in a thermal condensate state is
\be
 \braket{V_t}_{\sigma,\bar{\sigma};\beta} = \sum_J v_J \left( |\sigma_J(t)|^2 + \sinh^2{[\theta_J]}||t||^2 \right)
 \ee
where $||t||^2 = (t,t)_{L^2(\mathbb{R})}$ and,
\be \label{sigt}  \sigma_{J}(t) := \int_{\mathbb{R}} d\phi \; \overline{t}(\phi)\sigma_J(\phi) = (t , \sigma_J)_{L^2(\mathbb R)} \,. \ee
One can show that the quantity $\braket{V_t}_{\sigma,\bar{\sigma};\beta}$ can be expressed in terms of a non-local function of $\phi$, namely
\be \label{vnonloc}
 \braket{V_t}_{\sigma,\bar{\sigma};\beta} = \sum_J v_J \int_{\mathbb{R}^2} d\phi d\phi' \, t(\phi)\overline{t}(\phi') \braket{N_J(\phi,\phi')}_{\sigma,\bar{\sigma};\beta} 
\ee
where
\be N_J(\phi,\phi') := a_J^\dag(\phi)a_J(\phi') \ee 
is the off-diagonal number density (2-point) operator with expectation value,
\be 
 \braket{N_J(\phi,\phi')}_{\sigma,\bar{\sigma};\beta} =  \bar{\sigma}_J(\phi)\sigma_J(\phi') + \sinh^2[\theta_J]\delta(\phi-\phi') \,.
\ee
This generic non-locality of $t$-relational quantities with respect to $\phi$, for instance in \eqref{vnonloc}, is reasonable to expect simply as a technical feature that is characteristic of changing reference frames in general.

Lastly, the smearing functions $t(\phi)$ are understood as defining reference clock frames, the reasons for which will be made more clear in the next section. For now, we note that such a treatment is compatible with GFTs technically being background independent \emph{field} theories. Unlike in standard quantum field theory and particle physics, in any background independent field theory, a relational clock variable is expected to be defined as a genuine function, such as $t(\phi)$, rather than a coordinate on the base manifold, such as $\phi$. One common way to tackle this is to introduce an additional dynamical field into the system, deparametrize the full dynamics with respect to it, and use it as a clock, e.g. the Brown-Kuchar dust model in general relativity. In this case then, we notice that relational clocks are \emph{fields} over the spacetime base manifold. Hence, in GFTs one expects a relational clock to also be defined as a field over the base (Lie group) manifold. This is what we partially achieve in this work, by the use of functions $t(\phi)$. Having said that, we strictly refrain from assigning any further \emph{physical} interpretation to the function $t$, especially from the spacetime point of view, unlike the coordinate $\phi$ which has been motivated as a minimally coupled scalar matter field in previous works (see for instance \cite{Li:2017uao}).



\subsection{Relational functional dynamics} \label{eff2}

At this point, we come to the important task of expressing the effective GFT equations of motion in terms of the smearing functions. The goal is to arrive at a consistent dynamical description of the present system in a $t$-relational reference frame. Let us first reiterate our main line of reasoning. Smearing functions $t(\phi)$ are used in order to avoid divergences in the $\phi$-frame, e.g. in relational quantities like $\braket{V(\phi)}_{\sigma,\bar{\sigma};\beta}$. This leads to observables like $\braket{V_t}_{\sigma,\bar{\sigma};\beta}$. The condensate functional $\sigma_J(t)$ defined in \eqref{sigt}, then naturally takes on the role of the dynamical collective variable, instead of $\sigma_J(\phi)$. Therefore, the equations of motion \eqref{EqM} in terms of the variable $\phi$, must be rewritten suitably in terms of functions $t$, as follows.

We are seeking a differential equation of motion for $\sigma_J(t)$, encoding the same dynamics as \eqref{EqM}. We begin by noticing that the mass term in \eqref{EqM} can be written in terms of $\sigma_J(t)$ simply as
\be  M_J \sigma_J(t) = \int_\mathbb{R} d\phi \,  M_J \overline{t}(\phi) \sigma_J(\phi)  \ee
using the smearing. Therefore, as before, we see that smearing might offer us a way forward. We then smear the equations \eqref{EqM}, with an arbitrary square-integrable complex-valued smooth function $t(\phi)$, obtaining 
\be \label{smearEqM} \int_\mathbb{R} d\phi \, \overline{t}(\phi) \partial^2_\phi \sigma_J(\phi) - M_J \sigma_J(t) = 0 \,, \quad \forall J \,. \ee 
Now, in order to get a description completely in the $t$-frame, we require a suitable derivative operator with a well-defined action on functionals of $t$. For this, notice that,
\begin{align} \label{nt2new}
\int_\mathbb{R} d\phi \, \overline{t}(\phi) \partial^2_\phi \sigma_J(\phi) &= \left(- \int_\mathbb{R} d\phi \, \partial_\phi \overline{t} \; \frac{\delta}{\delta \overline{t}(\phi)} \right)^2 \sigma_{J}(t)  \nonumber \\
&=: \mathbf{d}_t^2 \sigma_{J}(t)
\end{align}
where we have used integration by parts, and the following boundary conditions for the smearing functions,
\be \label{bcond} 
\lim_{\phi \to \pm \infty}t(\phi) = 0 \; , \quad \lim_{\phi \to \pm \infty}\partial_\phi t(\phi) =0 \,. \ee
The operator $\mathbf{d}_t$ might seem to be a good choice for the functional derivative \cite{engel2013density,parr1994density}
that we are looking for. However, recall that we are working with complex-valued smearing functions. Thus, in our context, generic functionals of them depend on both $t$ and $\bar t$, which are considered to be independent variables, e.g. the norm $|\sigma_J(t)|^2 = \overline{\sigma_J(t)}\sigma_J(t)$ depends on two variables, $t$ and $\bar{t}$. The operator $\mathbf{d}_t$ must thus be extended by the conjugate term to obtain the hermitian differential operator
\be \nabla_t :=
- \int_\mathbb{R} d\phi \left( \partial_\phi \overline{t} \; \frac{\delta}{\delta \overline{t}(\phi)} +  \partial_\phi t \; \frac{\delta}{\delta t(\phi)} \right) . \ee 
Notice that, as required, we get an equation in terms of $\nabla_t$ that is analogous to \eqref{nt2new} above, i.e.
\be \label{nt2}
 \nabla_t^2 \sigma_{J}(t) = \int_\mathbb{R} d\phi \, \overline{t}(\phi) \partial^2_\phi \sigma_J(\phi) \,.
 \ee 
Therefore, the equations of motion \eqref{EqM} can be equivalently expressed as
\begin{align} \label{SEoM}
 \nabla_t^2 \sigma_{J}(t) - M_J \sigma_{J}(t) &= 0 \,, \quad \forall J
\end{align}  
for all square-integrable smooth functions $t(\phi)$ satisfying the boundary conditions \eqref{bcond}.

Note that if one was working with a dynamical model based on higher (than 2) order derivatives in $\phi$, or in general is interested in extending this setup to include arbitrary higher order generalisations of equation \eqref{nt2} above, then the boundary conditions \eqref{bcond} must be supplemented by vanishing of all higher order derivatives of $t$ in the limit $\phi \to \pm \infty$. In such a case then one could work with the space of Schwartz functions for instance, as the relevant set of smearing functions. However in the present analysis, we do not need to restrict to this special subspace of smooth functions, and the conditions \eqref{bcond} are sufficient. \

Few remarks are in order concerning the operator $\nabla_t$ and the associated $t$-relational setup. The operator $\nabla_t$ is a functional differential operator, consisting of functional derivatives with respect to $\bar{t}$ and $t$ \cite{engel2013density,parr1994density}. The flow induced by it is not on the GFT base manifold (in contrast with $\phi$), nor on a given spacetime, but rather on the space of smearing functions. Recalling that functional derivatives can be understood as generalisation of directional derivatives, then $\nabla_t$ essentially defines a flow with components along the directions of $(-\partial_\phi \overline{t})$ and $(-\partial_\phi t)$. Further, by construction this operator satisfies,
\be \label{opSAT} \nabla_t \ell(t) = \int_\mathbb{R} d\phi \, \overline{t}(\phi) \partial_\phi \ell(\phi)  \ee
for any function $\ell(\phi) \in L^2(\mathbb{R})$, where $\ell(\phi)$ satisfies \eqref{bcond}, and $\ell(t) := (t, \ell)_{L^2(\mathbb{R})}$. Notice that equation \eqref{opSAT} straightforwardly gives equation \eqref{nt2} used above.  The property in \eqref{opSAT} is important because it motivates the use of smearing functions as relational clock fields. As we have shown above, the $t$-relational dynamical quantities and equations are derived from an appropriate smearing of their (possibly non-local) $\phi$-relational counterparts. In particular, the $t$-functional equations of motion \eqref{SEoM} are simply the smearing of the $\phi$-dependent equations \eqref{EqM}. The interpretation of the smearing can then be clarified, as a first step, by considering a limiting case where the $t$-relational setup reduces to the usual $\phi$-relational one. Namely, if one takes a delta distribution\footnote{Note that a distribution would not satisfy the boundary conditions \eqref{bcond}, and also the operator $\nabla_t$ would not be well defined. However, this peculiar case is to be understood only as a limit, for instance by considering the limit of vanishing width for a family of Gaussian functions.} peaked on $\phi$, that is $t(\phi') = \delta(\phi' - \phi)$, then the full $t$-relational setup introduced above naturally reduces to the $\phi$-relational one that is used in all previous works in GFT cosmology. For instance, all the smeared quantities take their usual forms as functions of $\phi$, e.g. $\sigma_J(t) = \sigma(\phi)$, $a_J(t) = a_J(\phi)$, $\braket{N_J(t)}_{\sigma,\bar{\sigma};\beta} = \braket{N_J(\phi)}_{\sigma,\bar{\sigma};\beta}$. 

Along these lines, one can motivate specific choices of smooth clock functions peaked around points of the base manifold, namely values of $\phi$, for instance Gaussian functions. Such choices could then be interpreted as the implementation of a {\it deparametrization} procedure at the level of the background independent quantum theory. One could further understand the selection of a relational clock as a restriction to a special sector of physical states in the full (non-deparametrized) quantum theory, as was suggested in \cite{Kotecha:2018gof}. However, in general, one would expect to be able to realise such mechanisms in possibly different ways. For instance, in the present setting, this would correspond to a special choice of smearing functions $t$; while a different possibility is explored in \cite{dan}, in the context of zero temperature ($\beta=\infty$) GFT condensate cosmology. The complete details of mechanisms for deparametrization, how they relate to each other, and if there could be preferred choices, are interesting queries that are left for future investigations. In this article however, we proceed without any further restriction to a specific class of $t$ functions, and work with the general case. We note that the added generality may also allow for potential switching between relational reference frames in GFT, which is an expected feature of any background independent system devoid of an absolute notion of time or space (see for instance \cite{Hoehn:2018whn}, and references therein).

Furthermore, we notice that the $t$-relational setup presented here is constructed from the full non-deparametrized operator formulation of GFT, with the algebra satisfying \eqref{fullCRs}, and the deparametrization with respect to a relational clock field is implemented via introduction of smearing functions $t(\phi)$, as discussed above. Specifically, the kinematic description of the system is fully covariant, i.e.\ no preferred clock parameter $\phi$ (from possibly several ones \cite{Kotecha:2018gof}) or function $t(\phi)$ is chosen as \emph{the} clock. The dynamical description (equations \eqref{EoMCl}-\eqref{EoMEff}) is derived using the principle of least action, without the use of any relational Hamiltonian. This setup is then technically different from the one used in some recent works like \cite{Adjei:2017bfm,Wilson-Ewing:2018mrp}. The relational frame used in these other studies, as in all previous works in GFT cosmology \cite{Pithis:2019tvp,Oriti:2016acw,Gielen:2016dss}, is defined with respect to the parameter $\phi$, which as discussed above (see also \cite{ThGFT}) may lead to divergences. Also, the studies in \cite{Adjei:2017bfm,Wilson-Ewing:2018mrp} are based on a canonical quantization of already deparametrized classical GFT models. Specifically, the kinematic description is canonical with respect to a chosen clock variable $\phi$, with the algebra based on equal $\phi$-time commutation relations. Subsequently, the dynamical description is derived from a clock Hamiltonian. Having said that, the descriptions based on these two, a priori technically different, setups could eventually be related, since they encode the physics of a given system before and after deparametrization. This question is however tightly connected to the open issue of time in quantum gravity. The investigation of this possible relation may help in addressing the question of how physical time emerges in the present background independent theory of quantum gravity.

Returning to the equations of motion \eqref{SEoM}, let us use the standard polar decomposition
\begin{align}
 \sigma_{J }(t) = \zeta_J^t e^{i \eta_J^t}
\end{align}
where $\zeta_J^t = \sqrt{ \sigma_{J }(t)  \overline{\sigma_{J }(t)}}$ is the modulus and $\eta_J^t = \tan^{-1} \left( \frac{\text{Im} \, \sigma_{J }(t)}{ \text{Re} \, \sigma_{J }(t)} \right)$ is the phase of the condensate functional $\sigma_{J }(t) \in \mathbb{C}$. Note that the quantities $\zeta_J^t$ and $\eta_J^t$ do not correspond to the smearing of the modulus $\zeta_J(\phi)$ and the phase $\eta_J(\phi)$ of the condensate function $\sigma_{J }(\phi)$, which were used in the context of GFT cosmology in previous works \cite{Oriti:2016qtz}. Separating the real and imaginary parts of equations \eqref{SEoM}, we obtain
\begin{align}
\nabla_t^2 \zeta_J^t - \zeta_J^t (\nabla_t \eta_J^t)^2 - M_J \zeta_J^t = 0 \\
 2\nabla_t \zeta_J^t \nabla_t \eta_J^t + \zeta_J^t \nabla_t^2 \eta_J^t = 0
\end{align}
for all $J$. These two equations imply the existence of two constants of motion, as in the case of $\beta=\infty$ free theory \cite{Oriti:2016qtz}, given by
\begin{align} \label{ej}
 E_J & = (\nabla_t \zeta_J^t)^2 + (\zeta_J^t)^2 (\nabla_t \eta_J^t)^2 - M_J (\zeta_J^t)^2\\
 Q_J & = (\zeta_J^t)^2 \nabla_t \eta_J^t \label{qj}
\end{align}
satisfying $\nabla_t E_J = 0$ and $\nabla_t Q_J = 0$. 


\subsection{Effective dynamics for homogeneous and isotropic cosmology} \label{eff3}

Now, our investigation is based on four ingredients: the choice of quantum states in the full theory (here, the class of coherent thermal states based on the chosen Gibbs state), the choice of dynamics, the choice of relational observables, and finally the choice of a subclass of condensate wave functions. We have addressed the first three points in sections \ref{vgibbs}-\ref{eff2}, which brings us to the last one, which we address as follows. A notion of homogeneity in the present non-spatiotemporal background independent setting resides in: $(i)$ the use of a coherent condensate as the relevant phase for studying the effective cosmology extracted from a GFT model, and $(ii)$ having an additional left diagonal symmetry on the condensate wavefunction $\sigma(hg_i,\phi) = \sigma(g_i,\phi) , \; \forall h \in SU(2)$. In other words, it resides in the facts that the collective dynamics is encoded in a left- and right-invariant single-particle wavefunction $\sigma$, which is also the order parameter of the condensate $\braket{a_J(t)}_{\sigma,\bar{\sigma};\beta} = \sigma_J(t)$, where now $J \equiv (\vec{j},\iota_L,\iota_R)$; and that each $a$-quantum in the condensate is being described by the same wavefunction $\sigma$. Further, a notion of isotropy is implemented by fixing the spins at each vertex to be equal, fixing the two intertwiners to be equal (the geometric interpretation of which remains to be understood), and choosing a special class of intertwiners, namely the eigenvectors of the volume operator with the highest eigenvalue. We refer to past works for detailed discussions on these aspects, for instance \cite{Pithis:2019tvp,Oriti:2016qtz,Oriti:2016acw,Gielen:2014ila,Gielen:2013kla,Gielen:2013naa}. \

These restrictions imply that the condensate function is entirely determined by the value of a single spin $j$. It follows that the equations of motion \eqref{SEoM} reduce to one equation for each value of the $SU(2)$ spin label $j$,
\begin{align}  \label{EqMj}
 \nabla_{t}^2\sigma_{j}(t) - M_j \sigma_{j}(t) = 0 \,, \quad \forall j\in \mathbb{N}/2 \,.
\end{align} 
Consequently we have, $\forall j\in \mathbb{N}/2$
\begin{align}
\quad \nabla_t^2 \zeta_j^t - \zeta_j^t (\nabla_t \eta_j^t)^2 - M_j \zeta_j^t = 0 \label{CofM1}\\
 2\nabla_t \zeta_j^t \nabla_t \eta_j^t + \zeta_j^t \nabla_t^2 \eta_j^t = 0 \label{CofM2}
\end{align}
with the same conserved charges \eqref{ej} and \eqref{qj}, now labelled by the spin $j$.  \

Having set all the ingredients for a dynamical analysis, we can now proceed with the derivation of the effective dynamical equations for the average volume $\langle V_t \rangle$ in a coherent thermal state of the form \eqref{cts}, which include geometric volume fluctuations as discussed in \ref{vgibbs}. For simplicity of notation, we will drop the label $t$ on relational quantities (like $\zeta$, $\eta$ and volume averages) in the following. Relational volume average is given by
\begin{align}\label{Vt}
 {\bf V}:=\langle V_t \rangle = \sum_j v_j (\zeta_j^2 + s_j^2||t||^2)
\end{align}
where $s_j := \sinh\left[ \theta_{j}(\beta)\right]$. Using the effective equations of motion \eqref{CofM1} and \eqref{CofM2}, and the expressions for the constants of motion \eqref{ej} and \eqref{qj}, we obtain
\begin{align}
 {\bf V}' &:=  \nabla_t {\bf V} \nonumber \\ 
 &= \, 2 \sum \limits_j v_j \zeta_j \nabla_t \zeta_j \nonumber \\  
 &=  \, 2 \sum \limits_j v_j \zeta_j\ {\rm sgn}(\zeta'_j)\sqrt{E_j - \frac{Q_j^2}{\zeta_j^2} + M_j \zeta_j^2} \label{dVt} \,, \\ \nonumber \\
\label{d2Vt}
 {\bf V}'' &:= \nabla_t^2 {\bf V} = 2 \sum \limits_j v_j (E_j + 2 M_j \zeta_j^2) \,,
\end{align}
where we have used $\nabla_t ||t||^2 = 0$. From here on we shall assume $||t||^2=1$ for convenience. \

Then, the effective generalised Friedmann equations, including both quantum gravitational and statistical volume corrections are
\begin{align} \label{dVt1}
 \left(\frac{{\bf V}'}{3{\bf V}}\right)^2 &=  \frac{4}{9}\left( \frac{ \sum_j v_j \zeta_j\ {\rm sgn}(\zeta'_j)\sqrt{E_j - \frac{Q_j^2}{\zeta_j^2} + M_j \zeta_j^2} }{  \sum_j v_j \zeta_j^2 + \sum_j v_j s_j^2 } \right)^2 , \\
 \frac{{\bf V}''}{{\bf V}} &= \frac{2 \sum_j v_j (E_j + 2 M_j \zeta_j^2)}{\sum_j v_j \zeta_j^2 + \sum_j v_j s_j^2} \,.\label{d2Vt1}
\end{align}
These equations represent the relational evolution for the volume associated with a foliation labeled by the function $t$. Compared to the analogous equations obtained in \cite{Oriti:2016qtz}, the main difference arises due to the expression \eqref{Vt} for the average volume where there appears an additional statistical contribution $s_j^2$, which as we have described above, originates directly from the quantum statistical mechanics of the underlying quantum gravity theory.


\subsubsection{Late times evolution} \label{lateev}

In the following, we will make use of the quantities below that are formally defined as number densities corresponding to the different parameters characterising the different phases of the system:
\begin{align}
n_\co(j) = \zeta_j^2 \; &, \quad n_E(j) = \frac{E_j}{M_j} \,, \\
n_\tth(j) = s_j^2  \; &, \quad n_Q(j) = \frac{Q_j}{\sqrt{M_j}} \,, 
\end{align}
where $n_\co$ and $n_\tth$ (equal to \eqref{bose}) are the actual number densities (thus are non-negative) of the condensate and thermal parts of the full system. Different physical regimes can then be described in terms of relative strengths of these parameters. \

The domain $ n_E(j), n_Q(j) \ll n_\co(j)$ is understood as a classical limit where the volume is large but curvature is small \cite{Oriti:2016qtz}. In this regime, we have
\begin{align} \label{sclimit}
 {\bf V}' &\approx 2 \sum \limits_j {\rm sgn}(\zeta'_j) \, v_j \sqrt{M_j} \, \zeta_j^2 \,, \\
 {\bf V}'' &\approx 4 \sum \limits_j v_j M_j \zeta_j^2 \,,
\end{align}
giving the corresponding generalised evolution equations,
\begin{align}
 \left(\frac{{\bf V}'}{3{\bf V}}\right)^2 &= \frac{4}{9} \left( \frac{  \sum_j {\rm sgn}(\zeta'_j) \, v_j \sqrt{M_j} \, \zeta_j^2 }{ \sum_j v_j \zeta_j^2 + \sum_j v_j s_j^2} \right)^2 , \\
 \frac{{\bf V}''}{{\bf V}} &= \frac{4 \sum_j v_j M_j  \zeta_j^2}{\sum_j v_j \zeta_j^2 + \sum_j v_j s_j^2} \;.
\end{align}

Note that the thermal contribution ${\bf V}_\tth := \sum_j v_j n_\tth(j)$ is invariant under variations in the time function $t$, that is $\nabla_t {\bf V}_\tth = 0$. Hence, if the full system evolves such that the condensate number density $n_\co(j)$ increases monotonously in time, then eventually we will reach the domain where the condensate part, ${\bf V}_\co := \sum_j v_j n_\co(j)$, dominates the thermal cloud, that is ${\bf V}_\co \gg {\bf V}_\tth $. Thus considering
\be \label{flrwlim} n_\co \gg n_Q, n_E, n_\tth  \ee
and
\begin{align}\label{ClassCond}
 \forall j,\quad {\rm sgn}(\zeta'_j)=\pm 1 \ ,\ M_j \equiv M=3 \pi G 
\end{align}
where $G$ is Newton's gravitational constant, we obtain
\begin{align}
 \left(\frac{{\bf V}'}{3{\bf V}}\right)^2 &= \frac{4 \pi G}{3}\\
 \frac{{\bf V}''}{{\bf V}} &= 12 \pi G 
\end{align}
which are known to be the relational Friedmann equations in general relativity, for spatially flat FLRW spacetime with a minimally coupled massless scalar field \cite{Oriti:2016qtz}. These are the equations of motion of the quantum gravity system in a thermal condensate phase where both \eqref{flrwlim} and \eqref{ClassCond} are valid.
Physically, this regime where the condition \eqref{flrwlim} is satisfied, i.e.~ the contribution coming from the condensate is dominant while the statistical fluctuations are subdominant, corresponds to a phase that effectively mimics a system in a zero temperature condensate\footnote{Note that there could also be a classical regime where the statistical fluctuations are not subdominant. In other words, statistical fluctuations may be important even in regimes where quantum fluctuations are negligible. In the present setting, this could correspond to the case when $n_\co \gg n_E, n_Q$ holds true, while the interplay between $n_\co$ and $n_\tth$ is still relevant.}. Consequently, as shown above, in this regime we simply get the zero temperature condensate cosmology obtained in previous works \cite{Oriti:2016qtz,Oriti:2016ueo}. In this sense, the use of zero temperature condensates like $\ket{\sigma}$ can be understood more generally as a thermal quantum gravity condensate being in a dynamical regime where the condensate dominates, $n_\co \gg n_\tth$.



\subsubsection{Early times evolution} \label{earlyev}

We can also look at the evolution equations \eqref{dVt1} and \eqref{d2Vt1} in a different phase, in particular where the thermal contributions and the quantum corrections become relevant. For consistency, the choice \eqref{ClassCond}, which recovers the classical limit giving the correct late time behaviour, is assumed. \

Observe that the expression \eqref{dVt} for ${\bf V}'$ admits roots. Namely, there exist solutions $\{\zeta_j^o\}_j$ such that
\begin{align}
 {\bf V}'=0 \,.
\end{align}
The solution is given explicitly in terms of $M$ and the constants of motion $E_j,Q_j$ as
\begin{align}
n_\co^{o}(j) &= -\frac{1}{2}n_E(j) + \sqrt{\frac{1}{4}n_E(j)^2 + n_Q(j)^2} \;, \quad  \forall j 
\end{align}
where we have ignored the negative solutions, since $n_\co$ is the number density of the condensate and must be non-negative in general. At this stationary point, the total volume is 
\be {\bf V}^o = \sum_j v_j (n_\co^{o}(j) + n_\tth(j)) \ee 
which is clearly non-zero due to the non-vanishing thermal contribution in the present finite $\beta$ case, even if the condensate part were to vanish. However, as it happens, even $n_\co^{o}$ does not vanish as long as $E \neq 0$. In particular, $n_\co^{o} \neq 0$ even if $Q = 0$. To see this, notice that if $Q = 0$, then $E < 0$, which is evident from their expressions in equations \eqref{ej}-\eqref{qj}, assuming a positive $M$ as required by the correct classical limit \eqref{ClassCond}. In this case then, $n_\co^{o} = |n_E|$. In the more general case with $Q \neq 0$, both positive and negative $E$ are allowed in principle, but in each case we again have $n_\co^{o} > 0$. Thus, the expectation value ${\bf V}$ does not vanish when ${\bf V}' = 0$, implying the existence of a non-vanishing minimum\footnote{It can be verified that this is indeed a minimum since ${\bf V}''|_{n_\co^{o}} > 0$.} ${\bf V}^o$ of ${\bf V}$ throughout the evolution. Physically, in the context of homogeneous and isotropic cosmology, this means that the singularity has been resolved, and that the effective evolution displays a bounce, with a non-zero minimum of the spatial volume. \

Further, this ensures a transition between two phases of the universe characterised by the sign ${\rm sgn}(\zeta'_j)$, describing a contracting universe (${\rm sgn}(\zeta'_j)<0$) and an expanding one (${\rm sgn}(\zeta'_j)>0$). Each of these phases behaves according to the general relativistic FLRW evolution in the classical and non-thermal limits \eqref{flrwlim}, that is, when $\zeta_j$ (equivalently the condensate contribution to the volume) becomes very large with respect to all the constants of motion and the thermal contributions. However, as expected, these two phases display a non-standard evolution in general, especially when close to the bounce. This is the regime where $\zeta_j$ is comparable in magnitude to the other quantities present in the model. This leads to the particularly important question about the presence of a phase of accelerated expansion and its magnitude. \

To address this question, we proceed with a simplified analysis, where we make the approximation  of selecting one spin mode \cite{Oriti:2016qtz, deCesare:2016rsf}, thus dropping the sum over all spins in the various expressions. In this case, the generalised equations of motion \eqref{dVt1} and \eqref{d2Vt1} reduce to
\begin{align} 
 \left(\frac{{\bf V}'}{{\bf V}}\right)^2 &= 4M + 4\left(\frac{E_j \zeta_j^2 - Q_j^2 - 2 M \zeta_j^2 s_j^2 - M s_j^4}{(\zeta_j^2 + s_j^2)^2}\right) \,, \label{v'} \\
 \frac{{\bf V}''}{{\bf V}} &= 4M + 2\left(\frac{E_j - 2 M s_j^2}{\zeta_j^2 + s_j^2} \right) \label{v''} \;.
\end{align}
Now, the magnitude of a phase of accelerated expansion can be estimated in terms of the number of e-folds \cite{deCesare:2016rsf} given by
\begin{align}
 {\cal N} := \frac{1}{3} \ln \left( \frac{{\bf V}_{\rm end}}{{\bf V}_{\rm beg}}\right) &= \frac{1}{3} \ln \left( \frac{n_\co^{\rm end} + n_\tth}{n_\co^{\rm beg} + n_\tth}\right)
\end{align}
where ${\bf V}_{\rm beg}$ and ${\bf V}_{\rm end}$ are the average total volumes at the beginning and end of the phase of accelerated expansion respectively, and in terms of an acceleration \cite{deCesare:2016rsf} given by
\begin{align} 
\mfa := \frac{{\bf V}''}{{\bf V}} - \frac{5}{3}\left( \frac{{\bf V}'}{{\bf V}} \right)^2 \,.
\end{align} 
Using equations \eqref{v'} and \eqref{v''} above, we get
\begin{align} 
\mfa = & -\frac{8}{3}M + 2M \left(\frac{n_E - 2n_\tth }{n_\co + n_\tth}\right)\left(1 - \frac{10}{3}\frac{n_\co}{n_\co + n_\tth}\right) \nonumber \\ 
& + \frac{20}{3}M \left( \frac{n_Q^2 + n_\tth^2}{(n_\co + n_\tth)^2} \right)  \;.
\end{align}
Assuming that the bounce is the starting point of inflation, we have
\begin{align}
 n_\co^{{\rm beg}} = n_\co^o 
\end{align}
for any $j$. It can be checked straightforwardly that acceleration is positive at the beginning, i.e. $\mfa|_{\rm beg} > 0$ as required. Now, the end of accelerated expansion is characterized by $\mfa|_{\rm end} = 0 \,$, 
which gives,\\

\begin{align} \label{zend}
n_\co^{\rm end} = &\frac{3}{4} n_\tth - \frac{7}{8} n_E \nonumber \\ &+ \sqrt{ \frac{49}{64} n_E^2 +\frac{5}{2}n_Q^2 + \frac{9}{16} \left(n_\tth^2 -  n_E n_\tth \right)} 
\end{align}
assuming an expanding phase of the universe, i.e. ${\bf V}_{\rm end} > {\bf V}_{\rm beg}$, and non-negativity of $n_\co^{\rm end}$ even when $n_\tth$ is negligible. The number of e-folds can thus be estimated by,
\begin{align}
&e^{3\mcN} =  \nonumber \\
& \;\;\; \frac{ \frac{7}{4}\left(n_\tth - \frac{1}{2}n_E\right) + \sqrt{\frac{49}{64}n_E^2 + \frac{5}{2}n_Q^2 + \frac{9}{16}\left( n_\tth^2 - n_En_\tth \right) } }{ \left( n_\tth - \frac{1}{2}n_E \right) + \sqrt{\frac{1}{4}n_E^2 + n_Q^2 } } \,.
\end{align}

Since the quantities $n_E$, $n_Q$ and $n_\tth$ are all independent, we can observe three interesting regimes, giving approximate numerical values for $\mcN$:
\begin{align}
n_\tth, n_Q \ll n_E \ &:\qquad {\cal N} \approx \frac{1}{3} \ln\left( \frac{7}{4} \right) \approx 0.186\\
n_\tth , n_E \ll n_Q \ &:\qquad {\cal N} \approx \frac{1}{6} \ln\left(\frac{5}{2} \right) \approx 0.152\\
n_E , n_Q \ll n_\tth \ &:\qquad {\cal N} \approx \frac{1}{3} \ln\left( \frac{5}{2} \right) \approx 0.305
\end{align}
The upper bound on $\cal N$ in the previous zero temperature free theory analysis \cite{deCesare:2016rsf} is $0.186$, while here for finite $\beta$ free theory it is $0.305$, achieved in an early time limit. This difference is attributed to the only new aspect that we have introduced in the model, the thermal cloud of quanta of geometry. This shows that the number of e-folds can be increased, even without a non-linear dynamics. This fact is in contrast with the previous conclusions \cite{deCesare:2016rsf} that non-linear interaction terms in the GFT action are necessary to increase $\mcN$ (in which case, the interaction terms are naturally accompanied by their corresponding coupling constants, which are free parameters that can be fine-tuned to essentially give the desired value for $\mcN$). However, it remains that the increase in $\cal N$ in our case is very minimal and still not sufficient to match the physical observations, estimated at $\mcN \sim 60$. Nevertheless, this may be overcome if one goes beyond a static thermal cloud. In other words, a dynamical thermal cloud, which would be expected to be left over from a geometrogenesis phase transition (of an originally unbroken, pre-geometric phase), could have the potential to provide a viable mechanism for an extended phase of geometric inflation. The implementation of dynamical statistical fluctuations would require special care in order to avoid pathological behaviours with regards to the use of relational clock functions $t(\phi)$, in addition to the standard requirement of having stability via sufficiently subdominant or even decaying fluctuations in the relevant observables of the system, throughout the effective evolution of the universe. We do not treat this case in the present article and leave it for future works.

Finally, we note that the approximation of restricting to a single spin mode in the calculation above, does not alter the qualitative conclusion that a static thermal cloud is not sufficient to generate a satisfactory number of e-folds to match the observational estimate. However, a relaxation of the condition \eqref{ClassCond} for ${\rm sgn}(\zeta'_j)$, by considering a non-homogeneous distribution of the sign with respect to the modes $j$ while preserving the classical limit manifest in the emergence of Friedmann equations at large volumes, might give rise to a larger ratio between the volumes at the end and at the beginning of inflation and consequently a larger number of e-folds.


\section{Discussion and Outlook} \label{disc}

In this work we have studied some implications of the presence of statistical fluctuations in the context of group field theory by using coherent thermal states for condensate cosmology. We have modelled the quantum gravitational phase of the universe as a thermal condensate consisting of a condensate part representing an effective macroscopic homogeneous spacetime, and a static thermal cloud representing quantum geometric statistical fluctuations over it. This work provides the first steps toward building a GFT thermal condensate cosmology.

The model that we have presented recovers the expected cosmological dynamics at late times (when the thermal part is dominated by the condensate), but displays differences, with respect to earlier works with pure non-thermal condensates, at early times, when the thermal cloud dominates the condensate and in presence of quantum corrections. In particular, we have shown that the singularity is generically resolved with a bounce between a contracting and an expanding phase of the universe, and that there exists an early phase of accelerated expansion with an increased number of e-folds compared to those achieved in previous zero temperature analysis of free GFTs. This increase in the number of e-folds, obtained in the absence of interactions, is attributed to the presence of the thermal cloud. \

For our analysis, we have introduced an appropriate generalisation of the relational clock frame in GFT, by considering clock functions $t(\phi)$, implemented as smearing functions. Consequently, we have formulated the effective equations of motion and dynamical quantities as functionals of $t$. A more complete understanding of such relational frames in GFT and their precise relation to the concept of deparametrization is left to future work. \

Since the thermal cloud that we have considered in our study is static, considering a dynamical thermal cloud would be an important extension of this work, and would allow one to investigate further the dynamical implications of the presence of a thermal cloud. Another valuable extension would be to consider an interacting GFT model in the presence of thermal fluctuations, even with a static cloud. In fact, these aspects of having a dynamical thermal cloud and an overall interacting theory are intimately related, as would be expected. We discuss some of these issues below, and suggest further lines of investigations. \

There are three main features motivating our specific choice of state, namely a coherent thermal state of the form \eqref{cts} at inverse temperature $\beta$. Firstly, $\beta$ is assumed to be constant. Secondly, the state is such that the average number of quanta split neatly into a condensate and a non-condensate part, such that the zero temperature limit gives a pure $\beta$-independent condensate (see also equation \eqref{zerolim}). That is,
\begin{align}
\langle a^\dag  a \rangle_{\beta} &= n_\co + n_{\text{non-co}} \\
 \lim_{\beta \to \infty} \langle a^\dag a \rangle_{\beta} &= n_\co 
\end{align}
where in the present work the non-condensate part is taken to be thermal and at $\beta$-equilibrium. Having such a split is not only convenient in doing computations but also adds clarity to expressions in the subsequent analyses when considering the interplays between the two. Thirdly, the expectation value of the field operator in a coherent thermal state is temperature independent, i.e.
\be \label{orderpar} \braket{a_{J\alpha}}_{\sigma} = \sigma_{J\alpha}  = \braket{a_{J\alpha}}_{\sigma,\bar{\sigma};\beta}  \ee
thus being identical to the zero temperature case. Now at first sight this may seem contrary to our expectation that the condensate would be affected by the presence of a thermal cloud in general. But in fact this choice of state, satisfying the three properties noticed above, is entirely compatible with our approximation of neglecting interactions and using free dynamics. When temperature is switched on, quanta from the condensate are depleted into the thermal cloud. In a generally interacting case, both the thermal cloud and the condensate are interacting and dynamical by themselves, while also interacting with each other. However, in the present case of free dynamics, the thermal cloud will not interact with the condensate part, in addition to the quanta being free also within each part separately. Thus even though our state includes a thermal cloud, the coherent condensate part (described fully by its order parameter $\sigma$) will be unaffected by it, as depicted in equation \eqref{orderpar} above. This is further reasonable in light of constant $\beta$. If temperature were to change, 
say to increase, then we would expect more quanta to be depleted into the thermal cloud, and thus expect the state of the condensate to change. Since the condensate is characterised completely by the order parameter, it would also need to change. \

So overall, considering the class of states in which the order parameter is temperature independent, is a reasonable approximation when the temperature is constant and interactions are neglected. Excluding interactions ensures that the thermal cloud does not affect the condensate, while constant $\beta$ ensures that the amount of depletion is also constant, so together the condensate can indeed be approximated by a $\beta$-independent order parameter. In such a case we may be missing out on some interesting physics, however we take our case as a first step toward more advanced investigations in the future. \


From our discussion above, a physically important and interesting extension concerns the case of a dynamically changing $\beta$. In this case, the expected dominance of the condensate part over the thermal cloud at late times would not only be determined by a dynamically increasing condensate (as is the case in the present work), but also by what would be a dynamically decreasing temperature as the universe expands. \ 


Furthermore, one could extend the previous studies in GFT condensate cosmology \cite{Pithis:2019tvp,Oriti:2016acw,Gielen:2016dss} to the present setting with thermal fluctuations, and investigate various aspects including the dynamical analysis of quantum fluctuations, perturbations, anistropies and inhomogeneities. \

Finally, from a physical point of view, we presently understand $\beta$ as a statistical parameter that controls the extent of depletion of the condensate into the thermal cloud, and overall the strength of statistical fluctuations of observables in the system. The question remains whether it also admits a geometrical interpretation. Taking guidance from classical general relativity, we know that spatial volume generates a dynamical evolution in constant mean curvature foliations, wherein the temporal evolution is given by the so-called York time parameter. Constant York time slices are thus constant extrinsic curvature scalar (mean curvature) slices, and the two quantities are proportional to each other. In this case, one could attempt to understand $\beta$ as the periodicity in York time, equivalently in scalar extrinsic curvature (both of which are conjugates to the spatial volume). In particular for homogeneous and isotropic spacetimes, York time is further proportional to the Hubble parameter. A detailed investigation of such aspects and their implications would be interesting and is left for future work.

\bigskip

\begin{acknowledgments}
We are thankful to Daniele Oriti for many helpful discussions. M.A. acknowledges the support of the project BA 4966/1-1 of the German Research Foundation (DFG), and of the Polish National Science Center OPUS 15 Grant nr 2018/29/B/ST2/01250. I.K. is grateful to Benjamin Bahr and University of Hamburg for the generous hospitality during visits when this project was started and partially completed. I.K. acknowledges the support of the International Max Planck Research School for Mathematical and Physical Aspects of Gravitation, Cosmology and Quantum Field Theory, and the hospitality of the Visiting Graduate Fellowship program at Perimeter Institute where part of this work was carried out. Research at Perimeter Institute is supported in part by the Government of Canada through the Department of Innovation, Science and Economic Development Canada and by the Province of Ontario through the Ministry of Economic Development, Job Creation and Trade.
\end{acknowledgments}


\bibliographystyle{unsrt}
\bibliography{refvolcosmo}

\end{document}